\documentclass[conference]{IEEEtran}
\IEEEoverridecommandlockouts
\usepackage{cite}
\usepackage{amsmath,amssymb,amsfonts}
\usepackage{algorithmic}
\usepackage{graphicx}
\usepackage{textcomp}
\usepackage{xcolor}
\def\BibTeX{{\rm B\kern-.05em{\sc i\kern-.025em b}\kern-.08em
    T\kern-.1667em\lower.7ex\hbox{E}\kern-.125emX}}
\begin{document}

\title{Adaptive function approximation based on the Discrete Cosine Transform (DCT) \\
\thanks{This work was supported by MCIN/AEI/10.13039/501100011033 through the project IRENE (PID2020-115323RB-C31) and also by the Catalan government through the project SGR-Cat 2021-01207.}
}

\author{\IEEEauthorblockN{ Ana I. Perez-Neira}
\IEEEauthorblockA{\textit{SRCOM Unit} \\
\textit{CTTC - CERCA, UPC}\\
Castelldefels, Spain \\
ana.perez@cttc.es \\
ICREA Academia}
\and
\IEEEauthorblockN{Marc Martinez}
\IEEEauthorblockA{\textit{SRCOM Unit} \\
\textit{CTTC - CERCA}\\
Castelldefels, Spain \\
marc.martinez@cttc.es}
\and
\IEEEauthorblockN{Miguel A. Lagunas}
\IEEEauthorblockA{\textit{TSC Dept.} \\
\textit{UPC, CTTC}\\
Barcelona, Spain \\
m.a.lagunas@cttc.es}
}

\maketitle

\begin{abstract}
This paper studies the cosine as basis function for the approximation of univariate and continuous functions without memory. This work studies a supervised learning to obtain the approximation coefficients, instead of using the Discrete Cosine Transform (DCT). Due to the finite dynamics and orthogonality of the cosine basis functions, simple gradient algorithms, such as the Normalized Least Mean Squares (NLMS), can benefit from it and present a controlled and predictable convergence time and error misadjustment. Due to its simplicity, the proposed technique ranks as the best in terms of learning quality versus complexity, and it is presented as an attractive technique to be used in more complex supervised learning systems. Simulations illustrate the performance of the approach. This paper celebrates the 50th anniversary of the publication of the DCT by Nasir Ahmed in 1973.
\end{abstract}

\begin{IEEEkeywords}
DCT, LMS, supervised learning, function approximation.
\end{IEEEkeywords}

\section{Introduction}\label{s1}
The need for approximation of functions arises in many branches of applied mathematics and computer science. The different problems (regression, classification, estimation, ...) have received a unified treatment in statistical learning theory, where they are considered as supervised learning problems with machine learning at its core, \cite{b1},\cite{b2}. The need of nonlinear models becomes apparent and our focus in this paper is through the so-called kernel Hilbert spaces, which allow to map the input variables to a new space where the originally nonlinear task is transformed into a linear one. As a consequence, they base the approximation on just inner product operations, \cite{b3},\cite{b4},\cite{b5}. This paper revisits some traditional techniques concerning Volterra polynomial series expansion (e.g., \cite{b6,b7}) in order to propose a simpler and effective alternative. In contrast to the potential of neural networks, a common characteristic of this type of traditional models is that the kernel functions are pre-selected and they are fixed independently of the data. Although the price to pay is that the approximation error depends on the number of kernel functions, \cite{b8}, we will comment that this is one of the benefits of the proposed technique due to its energy compression properties. 

In this work we opt for trigonometric basis functions or Fourier representation. A continuous and univariate function $y=f(x)$ is considered, which must be learned from its noiseless samples ($y(n),x(n)$). Note that we do not aim to solve the multivariate case, which is more intrincate. Any multivariate continuous function can be approximated arbitrarily closely by a possibly infinite sum of harmonic functions, \cite{b9}. However, the curse of dimensionality must then be addressed and this is topic of current research (e.g., based on the random Fourier feature method, \cite{b10}). The purpose of this paper is to study the goodness of the adaptive Fourier approximation for the univariate case and show the potential of its use in future more involved systems that can build on it. Specifically, we study the cosine as basis function. In this way we can benefit from a better capacity for energy compaction than with Fourier basis (i.e., cosines and sines). In other words, with fewer basis functions we can concentrate most of the information of the function. The principle is the same as that of the Discrete Cosine Transform (DCT) versus the Discrete Fourier Transform (DFT), \cite{b11}. The DCT has many interesting applications (e.g., see the recent publications, \cite{b17}, \cite{b18} and \cite{b19}). The DCT was devised by Nasir Ahmed in 1973, thus celebrating now its 50th anniversary. We refer to \cite{b12} for a nice review of the impact that the DCT has in our “image society”.

The contributions of this paper are:
\begin{itemize}
  \item Innovative use of the DCT-based function approximation in a supervised learning setting.
  \item Due to the finite dynamics and orthogonality of the cosine basis functions simple gradient algorithms can benefit from it, and present a controlled and predictable convergence time and error misadjustment.
\end{itemize}

The remainder of the paper is organized as follows. Section \ref{s2} revisits the Volterra model under the Fourier transform perspective. Section \ref{s3} applies the gradient descent algorithms to obtain the approximation weights. Section \ref{s4} evaluates the studied approximation with some examples. Finally, Section \ref{s5} concludes the paper.

\emph{Notation:} Throughout the paper, a capital bold letter such as $\mathbf{A}$ represents a matrix, and a lower case bold letter $\mathbf{a}$ represents a vector. $\mathbf{A}^T$ is the transpose of a matrix, and $\mathbf{I}$ is the identity matrix.

\section{Fourier models in nonlinear functions}\label{s2}
The most popular model of a nonlinear system $y(x)$, called Volterra's, can be seen as a functional expansion over the exponential of an inverse Fourier transform
\begin{equation}
y(x)=\frac{1}{2\pi}\int Y(w) \,\exp{\left(jwx\right)}\,dw.\label{eq1}
\end{equation}
By developing the exponential in a Taylor series, the so-called Volterra model is obtained in which the base functions are powers of the input variable $x$; thus, it is a polynomial approximation
\begin{equation}
\exp{\left(jwx\right)}=\sum_{n=0}^{\infty} \frac{\left(jw\right)^n}{n!}\, x^{n}.\label{eq2}
\end{equation}
Using this expression in the inverse Fourier transform, we obtain \eqref{eq3} which is the polynomial approximation to the original function $y(x)$ when the order of Taylor series expansion is limited
\begin{equation}
y(x)=\sum_{n=0}^{\infty} \left(\frac{1}{2\pi}\frac{1}{n!}\int Y(w) \left(jw\right)^{n}\,dw\right)\, x^{n}=\sum_{n=0}^{\infty}c(n)x^{n}.\label{eq3}
\end{equation}
Note that the coefficients of the polynomial are basically the $n$-th derivative at the origin of the function divided by the factorial of $n$. 

The Volterra approach, despite its popularity, has not become widely used basically for two reasons. The first is that the base functions have a very large effective dynamic, $x^n$ in other words, they are not bounded unless the input dynamic is fixed. In any case, its effective dynamic range is very large and outside the assigned dynamics the function grows in an unbounded way. This is presented as an inconvenience when it comes to quantize this base functions or amplifying them, in the event that they have to be transmitted. The second reason, which is a consequence of the above, is that their dynamic variations make them risky functions for all gradient algorithms in learning with a reference or set of prototypes. This paper will come to this latter aspect in Section \ref{s3}. 

Another approximation, which, in principle, solves the problems mentioned in the previous paragraph, is the representation with a finite number of terms of the Fourier integral in \eqref{eq1}. Basically, as it is well-known, this approximation for a dynamic within $[$-1,1$]$ consists of assuming periodic the function to be approximated, of period equal to 2. Another benefit of this approach is that now the basis functions are orthogonal and this guarantees that the approximation error decreases according to the energy of the coefficient that are no longer used. This approximation can be carried out with the Discrete Fourier Transform (DFT) and retaining the coefficients of largest energy for its approximation.

\begin{figure}[htbp]
\centerline{\includegraphics[scale=0.38]{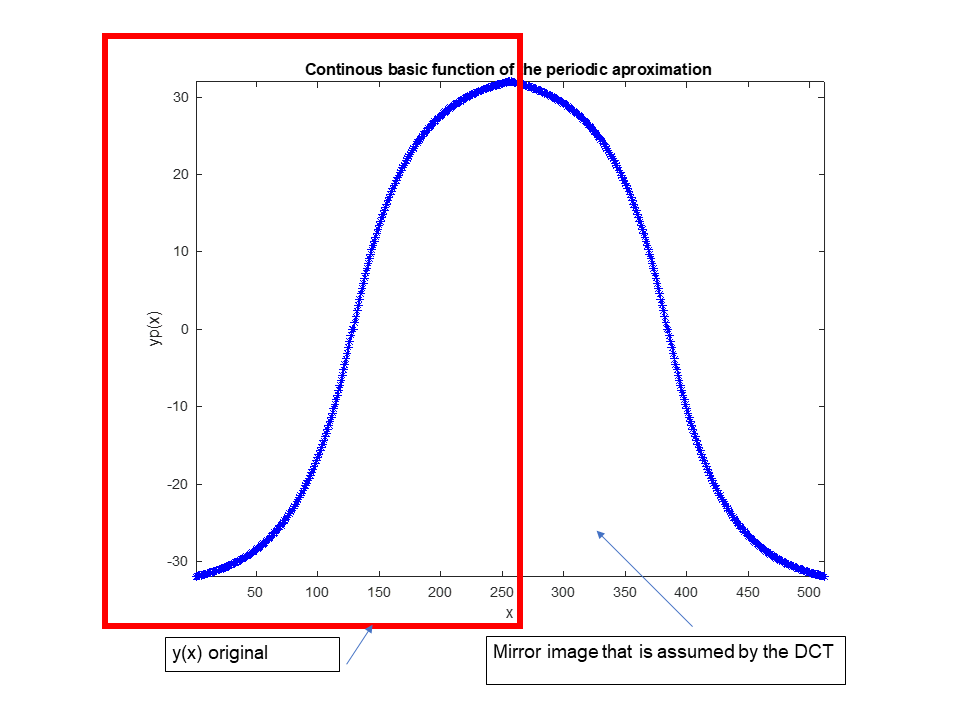}}
\caption{Mirror image that is assumed by the DCT in order to represent $y(x)$ in exchange for doubling the length of its periodicity.}
\label{fig1}
\end{figure}

However, if the function $y(x)$ is odd, the periodic prolongation generates a discontinuity at the edges. This discontinuity causes a greater number of components to carry out a good approximation so as to alleviate the called Gibbs phenomenon or ripple approximation in discontinuities. The harmonics caused by a discontinuity of order $n$, limits the maximum harmonic roll-off to 6(n+1) dBs per octave. If $y(x)$ is continuous in the mentioned interval, then the discontinuities always appear at the edges, which motivates to the use of the so-called cosine transform or Discrete Cosine Transform (DCT). This transform basically consists in the DFT of the symmetrically extended function, with even symmetry. In this way, the derivatives at the edges (i.e., at -1 and 1) of the extended function have been smoothed. Fig. \ref{fig1} depicts an example of this symmetric extension. Using the cosine transform, the expression used in the following of this document is shown below:
\begin{multline}
    \widehat{y}(x)= \sum_{i(q)\in S} c(q)\, \cos\left(\frac{\pi}{2N}\left(i(q)-1\right)\left(2z-1\right)\right)\\
z=\left(x+1\right)\frac{N}{2}\quad x\in\left[-1,1\right],\label{eq4}
\end{multline}
which is an alternative to the Volterra model in \eqref{eq3}. $N$ is the number or points to approximate. To accommodate the dynamics of the input $x$ to the approximation of the DCT, the input is mapped so that it is between the values 1 and N; thus resulting $z$. $S$ is the set that contains the $Q$ harmonics selected to make the approximation.

Given the value of $N$, the design of the approximation begins by taking $N$ samples of the function $y(x)$. After computing the DCT of those samples, the most representative coefficients are selected with an energy criterion. If the indices of the selected harmonics are $S=\{i(1), i(2),..., i(Q)\}$ the DCT becomes the set of coefficients $[$c(1), c(2),..., c(Q)$]$. The power of the approximation error results in
\begin{equation}
    p_e=|y(x)-\widehat{y}(x)|^2=\frac{1}{2}\sum_{i(q)\notin S}^{N} |c(q)|^2.\label{eq4bis}
\end{equation}

Note that the error spreads among coefficients that do not have a direct impact on a specific range of $y(x)$. This is an additional advantage over Volterra, which is a Taylor approximation, and therefore, presents better approximation accuracy near the origin of the function.

Instead of computing the DCT, this paper tries the adaptive design of the coefficients c(q) via supervised learning, through a set of pseudo random $n_{lea}$ training pairs $(y,x)$. The number of required coefficients must be initially determined or else estimated by excess and then remove the small coefficients in absolute value.If the function is odd, the coefficients will be even numbers. For example, for $Q$=9, the coefficients will be in [2:2:18]. The learning rule will be by instantaneous gradient of the mean square error as Section \ref{s3} explains.

\begin{figure}[htbp]
\centerline{\includegraphics{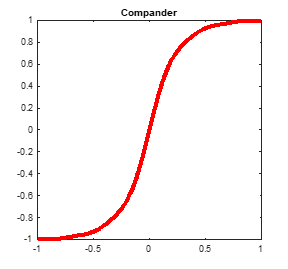}}
\caption{DCT approximation with 6 coefficients (Q=12) of a logarithmic compander.}
\label{fig2}
\end{figure}

The effect of smoothing the edges in the $x$-domain causes the number of coefficients to be greatly reduced in the DCT with respect to the Discrete Fourier Transform (DFT). Fig. \ref{fig2} shows the DCT approximation with 6 coefficients of a logarithmic compander. At the same time, with the cosine transform the base function for the development are not only orthogonal, but real and bounded, which is a significant advantage over its competitors as we show in the next section.

\section{Gradient Algorithms}\label{s3}
Gradient methods for the minimization of the mean squared error (MSE) in supervised learning are well-known, \cite{b13}. Next, we formulate the MSE in our approximation problem. First, we define the vector formed by the cosines of the harmonics selected for $x(n)$, where $n$ ($n$=1,...,$n_{lea}$) is the index of the samples for the learning process. The corresponding $z(n)$ is included in the expression \eqref{eq5}. $\mathbf{x}(n)$ and $\mathbf{i}$ are two column vectors of $Q$ components each.
\begin{equation}
    \mathbf{x}(n)=\cos\left(\frac{\pi}{2N}\left(\mathbf{i}-\mathbf{1}\right)\left(2z(n)-1\right)\right).\label{eq5}
\end{equation}
In this way, the output, $y_{sol}(n)$, would be \eqref{eq6}, being $\mathbf{c}$ the vector that contains the $Q$ coefficients to design
\begin{equation}
    y_{sol}(n)=\mathbf{c}^{T}\mathbf{x}(n).\label{eq6}
\end{equation}
The error is the difference between $y(n)$ and the output of the system expressed in \eqref{eq7}
\begin{equation}
    error(n)=y(n)-y_{sol}(n)\quad \,MSE=E\left(error(n)^{2}\right).\label{eq7}
\end{equation}
Likewise, the MSE is expressed, developing it further \eqref{eq8} results, where $\mathbf{R}$ is the covariance matrix of the vector $\mathbf{x}(n)$ in \eqref{eq5}, and the vector $\mathbf{p}$ is the correlation of the vector $\mathbf{x}(n)$ with the reference or $y(n)$ in \eqref{eq4}.
\begin{multline}
    MSE=p_{y}+E\left(\mathbf{c}^{T}\mathbf{x}(n)\mathbf{x}(n)^{T}\mathbf{c}\right)-\mathbf{c}^{T}\mathbf{p}-\mathbf{p}^{T}\mathbf{c}\\
    =p_{y}+\mathbf{c}^{T}\mathbf{R}\mathbf{c}-\mathbf{c}^{T}\mathbf{p}-\mathbf{p}^{T}\mathbf{c},\\ \mathbf{p}=E\{y(n)\mathbf{x}(n)\}\\ \mathbf{R}=E\{\mathbf{x}(n)\mathbf{x}(n)^T\}.\label{eq8}
\end{multline}
$p_{y}$ is the power of $y(n)$. It is worth noting the special properties of the vector $\mathbf{x}(n)$. First, its components are orthogonal, therefore the matrix $\bold{R}$ is diagonal. Second, the values on its diagonal will all be equal to $\frac{1}{2}$, corresponding to the power of a cosine. All this gives it a great advantage over using the polynomial model, where the dynamics of the basis functions increase exponentially and are not easy to controlled.

The gradient of the MSE is \eqref{eq9} 
\begin{equation}
    \nabla=\mathbf{R}\mathbf{c}-\mathbf{p}= 0.5\mathbf{c}-\mathbf{p}, \label{eq9}
\end{equation}
and the optimal coefficients, $\mathbf{c}^{*}$ are those that null this gradient
\begin{equation}
    \mathbf{c}^{*}= 2\mathbf{p}. \label{eq10}
\end{equation}
Note that, in fact, the vector $\mathbf{p}$ contains twice the first $Q$ coefficients of the function approximation in \eqref{eq4}.

By substituting the solution \eqref{eq10} into the MSE, the minimum error power is obtained
\begin{equation}
    MSE_{min}= p_{y}-2\mathbf{p}^T\mathbf{p}. \label{eq11}
\end{equation}
Remarkably, this minimum error power is equal to $p_e$ in \eqref{eq4bis}, that is the power of those DCT coefficients that have not been included in the approximation set $S$: 
\begin{equation}
    MSE_{min}= \frac{1}{2}\sum_{i(q)\neq S}^{N} |c(q)|^2. \label{eq12}
\end{equation}
The gradient descent algorithm consists in the following update recursion to obtain the coefficients:
\begin{equation}
    \mathbf{c}(t+1)= \mathbf{c}(t)-\mu \nabla(\mathbf{c}(t)), \label{eq13}
\end{equation}
where $\mu$ is the adaptation step, whose value must not exceed the quotient between 2 and the maximum eigenvalue of $\mathbf{R}$. If we take the trace of the matrix as an upper bound to this eigenvalue, then the adaptation step becomes bounded as in \eqref{eq14}
\begin{equation}
    \mu < \frac{4\alpha}{Q}=\frac{2\alpha}{0.5Q}=\frac{2\alpha}{Trace \mathbf{R}}< \frac{2\alpha}{\lambda_{max}}.\label{eq14}
\end{equation}

In practice, $\mu$ is chosen to fulfil the most tight upper bound $\frac{4\alpha}{Q}$. The parameter $\alpha$ influences the convergence time. If we denote by $t_{c}$ as the number of samples that the weakest mode (i.e., the minimum eigenvalue of $\mathbf{R}$, $\lambda_{min}$) takes to converge to a value of 0.01 from an initial value of 1, we get \eqref{eq15}.
\begin{multline}
    \left(1-\mu \lambda_{min}\right)^{t_c}=0.01  \rightarrow \\
   t_{c}^{theoretical} =-\frac{ln(0.01)}{\lambda_{min}\mu}=\frac{4.6\lambda_{max}}{2\alpha\lambda_{min}}=\frac{2.3}{\alpha}.\label{eq15}
\end{multline}

In practice we would like to account for the imperfections of the estimation of $\mathbf{R}$ and use the more tight upper bound of $\mu$ in \eqref{eq14}; thus, resulting \eqref{eq15bis}
\begin{multline}
    \left(1-\mu \lambda_{min}\right)^{t_c}=0.01  \rightarrow \\
   t_{c} =-\frac{log(0.01)}{\lambda_{min}\mu}<\frac{4.6 Q}{4\alpha \lambda_{min}}=\frac{4.6Q}{2\alpha}=\frac{2.3 Q}{\alpha}.\label{eq15bis}
\end{multline}

The technique in \eqref{eq13} is the so-called stochastic gradient descent algorithm, which converges in $t_c$ iterations to the optimal $MSE_{min}$ solution provided that $\mu$ satisfies the convergence condition in \eqref{eq14}. Once the algorithm has converged, it "locks" at the obtained solution. A very popular and simple implementation of it is the Normalized Least Mean Square Error (NLMS) algorithm, \cite{b14,b15}, which uses in the coefficients recursion \eqref{eq14} the instantaneous gradient \eqref{eq16} instead of the stochastic one
\begin{equation}
    \nabla_{ins}=-error(n)\bold{x}(n),\label{eq16}
\end{equation}
and the iteration index $t$ in \eqref{eq13} becomes the sample index $n$. In other words, it is an online learning algorithm where every new coefficient update is done with every new training pair ($y(n),x(n)$). The tighter upper bound in \eqref{eq14} is used for $\mu$. Now the approximation coefficients are a random variable whose mean is the optimal value in \eqref{eq10} and whose covariance matrix can be approximated as \eqref{eq17} (see \cite{b13} or chap.5 of \cite{b5}) 
\begin{equation}
    \bold{C}_{c}\approx\frac{\mu}{2}MSE_{min}\bold{I}.\label{eq17}
\end{equation}
The resulting final function approximation MSE, $MSE_{F}$, can be computed as
\begin{equation}
    MSE_{F}=MSE_{min}+Trace{\bold{C}_c \bold{R}}.\label{eq18}
\end{equation}

Therefore, the mismatch or excess error due to learning becomes \eqref{eq19}, where the tighter bound of $\mu$ in \eqref{eq14} has been used
\begin{equation}
    \zeta=\frac{MSE_F-MSE_{min}}{MSE_{min}}=\frac{\mu}{2}Trace{\mathbf{R}}=\alpha.\label{eq19}
\end{equation}

We conclude that $\alpha$ impacts on both the excess error and the convergence time $t_c$. After setting $\alpha$ and $Q$, the performance of the proposed adaptive technique can be predicted. This is shown next.

\section{Evaluation}\label{s4}
In this section, we illustrate the quality of the adaptive design using the cosine transform to approximate some chosen odd functions. The number of training pairs (i.e., {$y(n),x(n)$}) is equal to 50,000. Since they are odd functions, the $\bold{i}$ vector contains only even numbers. The number of coefficients retained is such that the approximation error of the cosine transform in \eqref{eq4bis} is 0.01$\%$, which, depending on the function to approximate, means different number of coefficients, as we see next. This papers does not aim to present a comparative evaluation, since the NLMS algorithm is very well study in the literature. Our aim is to verify the performance that we have predicted in the previous section when the DCT basis functions are used.

\begin{figure}[htbp]
\centerline{\includegraphics[scale=1.2]{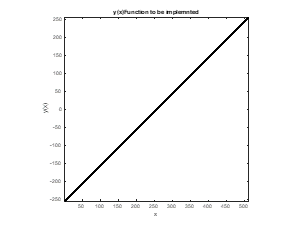}}
\caption{Linear function to be approximated.}
\label{fig3}
\end{figure}

\begin{figure}[htbp]
\centerline{\includegraphics[scale=1.2]{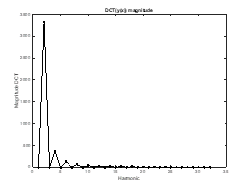}}
\caption{Optimal coefficients of the cosine transform. The DCT has been computed with $N$=512 points}
\label{fig4}
\end{figure}

\begin{figure}[htbp]
\centerline{\includegraphics[scale=1.5]{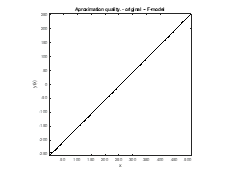}}
\caption{Linear function versus its approximation with the cosine transform. The approximation error is 0.01$\%$ because only 5 coefficients of \eqref{eq4} have been considered.}
\label{fig5}
\end{figure}

\begin{figure}[htbp]
\centerline{\includegraphics[scale=1.5]{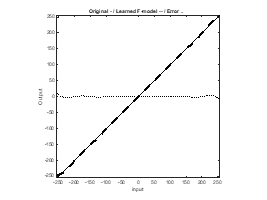}}
\caption{Linear function versus its NLMS approximation with $\alpha$=0.001 and 5 coefficients in dash lines. Their difference or error is plotted with dots}
\label{fig6}
\end{figure}

\begin{figure}[htbp]
\centerline{\includegraphics[scale=1.4]{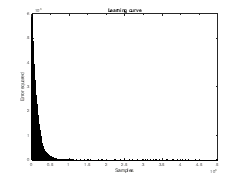}}
\caption{NLMS error curve with $\alpha=0.001$ and $Q$ = 5 coefficients.}
\label{fig7}
\end{figure}

\begin{figure}[htbp]
\centerline{\includegraphics[scale=1.5]{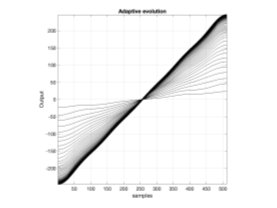}}
\caption{Evolution of the approximated curve during the NLMS learning.}
\label{fig8}
\end{figure}

Even though the approximation of a linear function may seem trivial, it might not be so if the basis functions are not. For this reason, first, we study the approximation of the linear function of Fig.\ref{fig3}, whose cosine transform is shown in Fig.\ref{fig4}. By retaining a 99.99$\%$ of the transform (i.e., 0.01$\%$ of DCT approximation error), we obtain that 5 coefficients are needed. Fig. \ref{fig5} superposes the original function with the approximation obtained with the 5 coefficients. Next we use the described NLMS with $\alpha$=0.001 to learn these coefficients. This cannot produce a better function approximation than the one in Fig. \ref{fig5}. Fig. \ref{fig6} shows the obtained approximation after convergence, superposed to the original function with dash lines. The final error is also plotted (i.e., $y(x)-y_{sol}(c(50,000))$. Fig. \ref{fig7} shows the error learning curve. The convergence time is shown in table \ref{table:1}. In this table, the measured $t_{c}$ is obtained when the error is 1$\%$ of the initial one. We observe that the conservative value that is given by \eqref{eq15bis} computes better the visual convergence shown by the curve. However, we note that both $t_{c}$ in \eqref{eq15} and \eqref{eq15bis} are computed such that $\lambda_{min}$ reduces its value a $99\%$, but it is not $100\%$ reduced. This is the reason why we have used 50,000 training pairs (i.e., NLMS iterations), such that we can measure a final error that is closer to the theoretical one of \eqref{eq18}. With this measured final error, $MSE_{F}=3.6804$, the misadjustment is computed and shown in table \ref{table:2}, together with the theoretical one. The minimum error, $MSE_{min}=3.67$, has been obtained by estimating the statistics in \eqref{eq11}. The power of the final error, $MSE_{F}$, has been computed with the average of the last 3,000 realizations. Finally, Fig. \ref{fig8} depicts the adaptive evolution of the approximated function. 

\begin{table}[h!]
\caption{Convergence results for the learning in fig.\ref{fig7}.}
\label{table:1}
\centering
\begin{tabular}{ |p{2.4cm}|p{2cm}|p{2cm}|}
 \hline
 \multicolumn{3}{|c|}{Convergence time} \\
 \hline
 $t_c^{theoretical}$ eq.\eqref{eq15}& measured $t_{c}$ & $t_{c}$ eq.\eqref{eq15bis}\\
 \hline
 2300   & 2543   & 11500\\
 \hline
\end{tabular}
\end{table}

\begin{table}[h!]
\caption{Misadjustment results for the learning in fig.\ref{fig7}.}
\label{table:2}
\centering
\begin{tabular}{ |p{3.5cm}|p{2cm}|}
 \hline
 \multicolumn{2}{|c|}{Misadjustment} \\
 \hline
 measured $\zeta$ & $\zeta$ eq.\eqref{eq19}\\
 \hline
 $100\frac{|3.6804-3.67|}{3.67}=0.28\%$ & 100$\alpha =0.1\%$\\
 \hline
\end{tabular}
\end{table}

\begin{figure}[htbp]
\centerline{\includegraphics[scale=1.2]{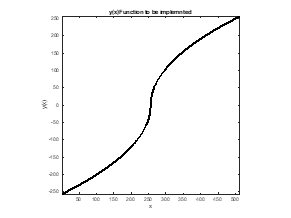}}
\caption{Square root function $\sqrt{x}$ to be approximated.}
\label{fig9}
\end{figure}

\begin{figure}[htbp]
\centerline{\includegraphics[scale=1.2]{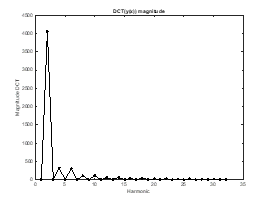}}
\caption{Optimal coefficients of the cosine transform. The DCT has been computed with $N$=512 points}
\label{fig10}
\end{figure}

\begin{figure}[htbp]
\centerline{\includegraphics[scale=1.2]{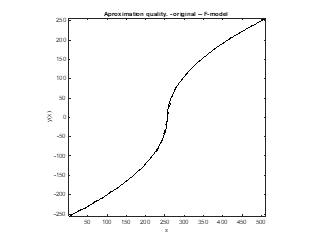}}
\caption{Square root function versus its approximation with the cosine transform. The approximation error is 0.01$\%$ because only 12 coefficients of \eqref{eq4} have been considered.}
\label{fig11}
\end{figure}

\begin{figure}[htbp]
\centerline{\includegraphics[scale=1.2]{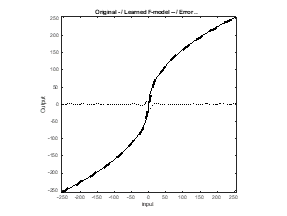}}
\caption{Linear function versus its NLMS approximation with $\alpha=0.001$ and 12 coefficients in dash lines. Their difference or error is plotted with dots}
\label{fig12}
\end{figure}

\begin{figure}[htbp]
\centerline{\includegraphics[scale=1.5]{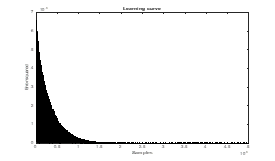}}
\caption{NLMS error curve with $\alpha=0.001$ and $Q$ = 12 coefficients.}
\label{fig13}
\end{figure}

\begin{figure}[htbp]
\centerline{\includegraphics[scale=1.5]{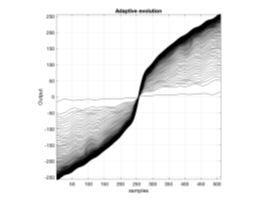}}
\caption{Evolution of the approximated curve during the NLMS learning..}
\label{fig14}
\end{figure}

Next, we examine another example: the square root function depicted in Fig. \ref{fig9}, whose cosine transform is shown in Fig.\ref{fig10}. By retaining a 99.99$\%$ of the transform (i.e., 0.01$\%$ of DCT approximation error), we obtain that 12 coefficients are needed. Fig. \ref{fig11} superposes the original function with the approximation obtained with the 12 coefficients. Now we use the described NLMS with $\alpha$=0.001 to learn these coefficients. This cannot produce a better function approximation than the one in Fig. \ref{fig11}.  Fig. \ref{fig12} shows the obtained approximation after convergence, superposed to the original function with dash lines. The final error is also plotted (i.e., $y(x)-y_{sol}(c(50,000))$. Fig. \ref{fig13} shows the error learning curve. The convergence time is shown in table \ref{table:3}. Again, as in Fig. \ref{fig8}, we observe that the conservative value that is given by \eqref{eq15bis} computes better the visual convergence shown by the curve. The computed and the theoretical misadjustments are shown in table \ref{table:4}. Finally, Fig.\ref{fig14} depicts the adaptive evolution of the approximated function.

\begin{table}[h!]
\caption{Convergence results for the learning in fig.\ref{fig13}.}
\label{table:3}
\centering
\begin{tabular}{ |p{2.4cm}|p{2cm}|p{2cm}|}
 \hline
 \multicolumn{3}{|c|}{Convergence time} \\
 \hline
 $t_c^{theoretical}$ eq.\eqref{eq15}& measured $t_{c}$ & $t_{c}$ eq.\eqref{eq15bis}\\
 \hline
 2300   & 2819   & 27600\\
 \hline
\end{tabular}
\end{table}

\begin{table}[h!]
\caption{Misadjustment results for the learning in fig.\ref{fig13}.}
\label{table:4}
\centering
\begin{tabular}{ |p{3.6cm}|p{2cm}|}
 \hline
 \multicolumn{2}{|c|}{Misadjustment} \\
 \hline
 measured $\zeta$ & $\zeta$ eq.\eqref{eq19}\\
 \hline
 $100\frac{|5.0902-5.0519|}{5.0519}=0.76\%$ & 100$\alpha =0.1\%$\\
 \hline
\end{tabular}
\end{table}

\begin{figure}[htbp]
\centerline{\includegraphics[scale=1.5]{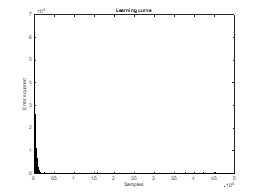}}
\caption{LMS error curve with $\alpha=0.01$ and $Q$ = 12 coefficients.}
\label{fig15}
\end{figure}

In order to illustrate the impact of $\alpha$ in the learning, we repeat the approximation of the function in Fig.\ref{fig9} with $\alpha$=0.01. As this parameter is higher now, the convergence time is reduced, as it is shown in the learning curve of Fig.\ref{fig15} and the convergence times of table \ref{table:5}. The corresponding misadjustments are in table \ref{table:6}. Interestingly, although the $MSE_{F}$ has increased with respect to the approximation with $\alpha$=0.001, as expected, the misadjustment is lower. The reason is that $MSE_{min}$ has also increased, because it is computed with \eqref{eq11} using the measured $p_{y}$ and $\mathbf{p}$, which are less accurately estimated than with $\alpha$=0.001. Therefore, as commented before, it is important to let the learning run much more iterations than those dictated by $t_{c}$, as it is crucial to obtain a good approximation and $t_{c}$ in \eqref{eq15} and \eqref{eq15bis} has been considered to converge to a value of 0.01, instead of zero, from an initial value of 1. 

\begin{table}[h!]
\caption{Convergence results for the learning in fig.\ref{fig15}.}
\label{table:5}
\centering
\begin{tabular}{ |p{2.4cm}|p{2cm}|p{2cm}|}
 \hline
 \multicolumn{3}{|c|}{Convergence time} \\
 \hline
 $t_c^{theoretical}$ eq.\eqref{eq15}& measured $t_{c}$ & $t_{c}$ eq.\eqref{eq15bis}\\
 \hline
 230   & 296   & 3552\\
 \hline
\end{tabular}
\end{table}

\begin{table}[h!]
\caption{Misadjustment results for the learning in fig.\ref{fig15}.}
\label{table:6}
\centering
\begin{tabular}{ |p{3.6cm}|p{2cm}|}
 \hline
 \multicolumn{2}{|c|}{Misadjustment} \\
 \hline
 measured $\zeta$ & $\zeta$ eq.\eqref{eq19}\\
 \hline
 $100\frac{|5.713-5.6879|}{5.6879}=0.44\%$ & 100$\alpha =1\%$\\
 \hline
\end{tabular}
\end{table}

\section{Conclusions}\label{s5}
The DCT approximation involves kernels or functions of finite dynamics, regardless of the input signal dynamics and are also orthogonal. With the aforementioned orthogonality, the use of error gradient methods, even those with the lowest complexity, have a deterministic and diagonal autocorrelation matrix of the data with null eigenvalue spread, i.e. all the eigenvalues are equal to ½. Due to this feature of the DCT model, the convergence time and the error misadjustment can be better controlled than with other basis function representations, such as Volterra. This makes the proposed supervised function approximation technique ranks among the best in terms of quality of the learning versus complexity. The new model and its capacity for easy learning and predictable quality, suggest its use in more complex supervised learning systems, even when the functions to approximate are not known. At the same time the cosine base functions can be used as transport waveforms and be directly radiated in over-the-air-computing systems as it is shown in \cite{b16}. Future work will address the use of the proposed technique to approximate high order cost functions that allow to separate source signals that are simultaneously received in an interference channels as in \cite{b17}.

\vspace{12pt}

\end{document}